\documentclass[twocolumn,showpacs,prl,preprintnumbers,amsmath,amssymb]{revtex4}
\date{\today}

\usepackage{graphicx}
\usepackage{dcolumn}
\usepackage{bm}
\usepackage{setspace}

\newcommand{\rr}{\mbox{\boldmath $r$}}
\begin{document}
\title{
Signature of Wave Chaos in Spectral Characteristics of Microcavity Lasers
}

\author{
Satoshi Sunada$^1$,
Susumu Shinohara$^2$,
Takehiro Fukushima$^3$,
and
Takahisa Harayama$^{4}$
}
\affiliation{
$^1$Faculty of Mechanical Engineering, Institute of Science and
Engineering, Kanazawa University, Kakuma-machi Kanazawa, Ishikawa 920-1192, 
Japan \\
$^2$NTT Communication Science Laboratories, NTT Corporation,
2-4 Hikaridai Seika-cho Soraku-gun, Kyoto 619-0237, Japan \\
$^3$Department of Information and Communication Engineering, 
Okayama Prefectural University, 111 Kuboki Soja, Okayama 719-1197,
Japan\\
$^4$Department of Applied Physics, School of Advanced Science and
Engineering, Waseda University, 3-4-1 Okubo, Shinjuku-ku, Tokyo
169-8555, Japan
}

\begin{abstract}
We report an experimental investigation on the spectra of fully chaotic
 and non-chaotic microcavity lasers under continuous-wave operating
 conditions.
It is found that fully chaotic microcavity lasers operate in single
 mode, whereas non-chaotic microcavity lasers operate in multimode. 
The suppression of multimode lasing for fully chaotic microcavity lasers
 is explained by large spatial overlaps of the resonance wave functions
 that spread throughout the two-dimensional cavity due to the ergodicity
 of chaotic ray orbits. 
\end{abstract}

\pacs{42.55.Sa, 05.45.Mt, 42.55.Px}
\maketitle
Various active devices ranging from musical instruments to lasers
generate oscillating states with well-defined frequencies from the
interplay between resonator geometry and an active nonlinear element 
\cite{Idogawa1993,Haken,Harayama2011}.  
Understanding and controlling the formation of such self-organized oscillating states 
is important in device physics and related applications. 
As a specific example, two-dimensional (2D) microcavity lasers have
attracted considerable attention over the past decades \cite{Nockel1997,Gmachl1998,Cao2015}.
Depending on the cavity shapes, they can exhibit a variety of lasing states 
through the interaction between 
the light field and active gain material \cite{Harayama2011}.   
The studies of 2D microcavity lasers have led to a wide range of applications,
including low-threshold microlasers with unidirectional emission 
\cite{Wiersig2008,Yan2009,Song2009,C-HYi2009,QJWang2010,Chern2003,Jiang2016}, 
low-coherence microlasers 
\cite{Redding2015}, and fast random signal generation \cite{Sunada2014}.
Moreover, 2D microcavity lasers have served as a platform 
for experimentally addressing fundamental issues such as quantum/wave
chaos in open systems \cite{Schwefel2004,Shinohara2007,SBLee2007,Hackenbroich1997,Shinohara2011,SBLee2002} and non-Hermitian physics 
\cite{Liertzer2012,Peng2014}.

Both experimental and theoretical studies on 2D microcavity lasers with
various cavity shapes have 
shown that the lasing emission patterns
can be well characterized by a limited number of low-loss resonance modes 
(i.e., eigenmodes with low-losses in passive cavities)
\cite{Harayama2011,Nockel1997,Gmachl1998}.   
These studies
 motivated research on the resonance characteristics of microcavities
\cite{Cao2015},
and presented an interesting problem on 
the mechanism of mode selection, namely, which modes and how
many of them can simultaneously lase for a given cavity shape and pumping condition.
The mode selection is mainly due to nonlinear
mode couplings (via gain materials) 
such as mode competition \cite{LaserPhysics,SargentIII1993}, i.e.,
lasing of some modes suppresses that of other modes.
While mode competition and its resulting lasing phenomena 
have been actively studied by a nonlinear dynamical model
\cite{Harayama2003,Sunada2005,Harayama2005,Harayama2003-2,Sunada2013} and
a steady state {\it ab initio} laser theory \cite{Hakan2006,LiGe2010},
the effect of the cavity shapes on the selection of lasing modes
is still unclear. 

In this Letter, we shed light on the effect of cavity
shapes on the number of lasing modes
by systematically investigating the lasing spectra of 2D semiconductor microcavity lasers 
under continuous-wave (cw) operating conditions.
We focus on two 2D microcavities that are categorized 
as fully chaotic and
non-chaotic cavities from a ray optics point of view
\cite{Harayama2011}, 
where all of the internal ray orbits are chaotic 
in the former and none of the ray orbits 
are chaotic in the latter.
The lasing in fully chaotic cavities has been studied to some extent
\cite{Redding2015,Audet2007,MWKim2012,Shinohara2008,Harayama2003_Scar,Fang2007,Lebental2006,Choi2008},
however, experimental investigations on the lasing spectra  
have not yet been performed in detail,
such as by changing the cavity sizes and degrees of
deformation from a circular cavity and comparing the spectra with
those of non-chaotic cavities.
Our investigations reveal that 
lasers with fully chaotic cavities (hereafter referred to as fully
chaotic cavity lasers) exhibit strong suppression of
multimode lasing and operate in single mode regardless of the cavity size 
and deformation parameters, 
whereas lasers with non-chaotic cavities (non-chaotic cavity lasers) operate in
multimode. 
This is the first experimental result that systematically reveals 
the relationship between cavity shapes and spectral characteristics, 
despite the fact that 
single mode emission in fully chaotic cavity lasers 
has been reported previously in the literature \cite{Sunada2013,Audet2007,MWKim2012}.

\begin{figure}[t]
\begin{center}
  \begin{tabular}{c}
\hspace*{-0.4cm}
\raisebox{0.0cm}{\includegraphics[width=8cm]{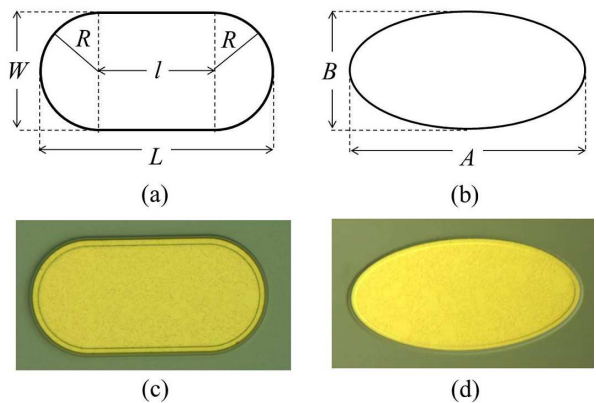}}
  \end{tabular}
\end{center}
\vspace{-4mm}
\caption{\label{fig:Shape} (Color online) (a) Stadium cavity and (b)
  elliptic cavity.  (c)(d) Optical microscope images of the fabricated
  lasers, where both cavities have an area $S$ = 8748 $\mu$m$^2$ and
  an aspect ratio $p=0.5$.
}
\end{figure}

In this work, we focus on a specific shape of the fully chaotic cavity  
called the stadium \cite{Bunimovich1979}, which 
is widely used for classical and quantum chaos studies.
As shown in Fig. \ref{fig:Shape}(a), 
the stadium cavity consists of two straight lines of length $l$ and two half
circles of radius $R$. We define the aspect ratio parameter $p=W/L$,
where $W=2R$ is the length of the minor axis and $L$ = $2R+l$ is the
length of the major axis. 
For the non-chaotic cavity, we focus on the elliptic cavity defined
in Fig. \ref{fig:Shape}(b). 
It is known that a closed elliptic cavity is an integrable system \cite{Berry1981}, 
thus exhibiting no chaotic behavior.
In the same manner as for the stadium cavity, we define the aspect
ratio parameter for the elliptic cavity as $p=B/A$, where $A$ and $B$
are the lengths of the major and minor axes, respectively.

We fabricated semiconductor microcavities with the stadium and
elliptic cavities by applying a reactive-ion-etching technique to a
graded index separate-confinement-heterostructure (GRIN-SCH) 
single-quantum-well GaAs/Al$_x$Ga$_{1-x}$As structure grown by MOCVD 
(See Ref. \cite{Fukushima2004} for details on the layer structures and
fabrication process).
The fabricated lasers are shown in Figs. \ref{fig:Shape}(c) and
\ref{fig:Shape}(d).
In our experiments, the lasers were soldered onto aluminum nitride
submounts at 20 $\pm$ $0.1$ $^{\circ}$C and electrically driven with cw
current injection.
The optical outputs were collected with anti-reflection-coated lenses
and coupled to a multimode optical fiber via a 30-dB optical isolator. 

\begin{figure}
\begin{center}
  \begin{tabular}{c}
\hspace*{-0.4cm}
\raisebox{0.0cm}{\includegraphics[width=8.8cm]{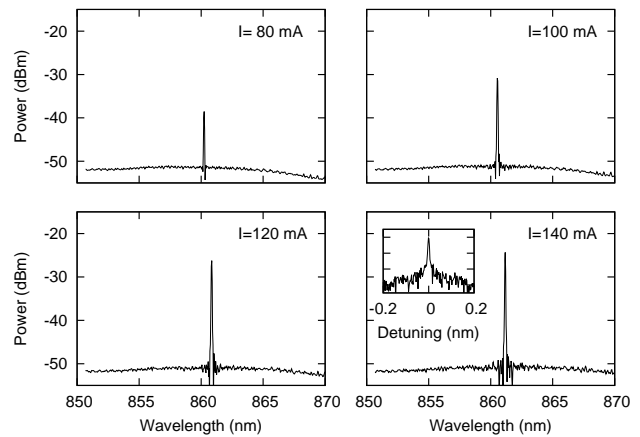}}
  \end{tabular}
\end{center}
\vspace{-4mm}
\caption{\label{fig:Stadium_Spec}
Spectra of the stadium cavity laser with a cavity area $S = 8748$
$\mu$m$^2$ and an aspect ratio $p = 0.5$ for various current values.  
The threshold current is approximately 74 mA. 
}
\end{figure}

Figure \ref{fig:Stadium_Spec} shows the lasing spectra for the
stadium cavity laser with a cavity area $S$= 8748 $\mu$m$^2$ and $p$ = 0.5
($R$ = 35 $\mu$m and $l$ = 70 $\mu$m)
for various current values above a threshold current $I_{th} \approx$ 74 mA. 
Because of the large area, the number of modes within the gain band 
was estimated to be a few thousand. 
Nevertheless, the lasing spectra exhibit only a single
sharp peak regardless of the current value $I$.  
The peak continuously red-shifts as $I$ increases, due to a thermal effect. 
As shown in the inset of Fig. \ref{fig:Stadium_Spec}, the linewidth of the peak 
is approximately 0.004 nm, which is close to the resolution limit (0.002 nm)
of our spectrum analyzer (Advantest Q8347). 
For further evidence on single-mode lasing (i.e., single-wavelength lasing), 
we measured the power spectrum of the output intensity 
using a photodetector with a bandwidth of 12.5 GHz and confirmed that  
there were no beat frequencies due to multimode lasing in the spectrum 
\cite{Sunada2013}. 
In addition, we evaluated the number of lasing modes  
by measuring the contrast of the far-field emission pattern 
\cite{Contrast}. 
Using a relationship between the contrast and the number of lasing modes \cite{Redding2015},
we obtained a result that suggested single-mode operation. 

\begin{figure}
\begin{center}
  \begin{tabular}{c}
\hspace*{-0.4cm}
\raisebox{0.0cm}{\includegraphics[width=8.8cm]{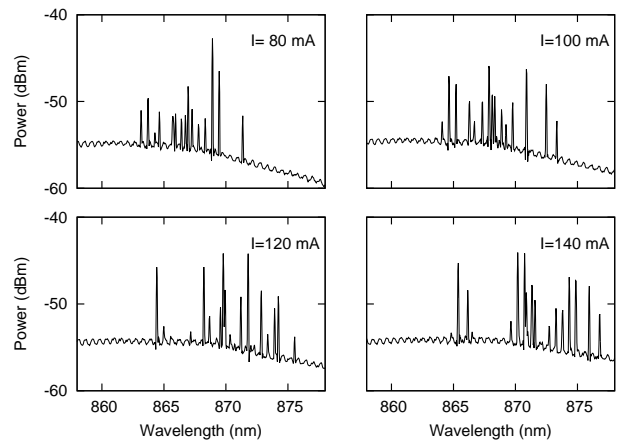}}
  \end{tabular}
\end{center}
\vspace{-4mm}
\caption{\label{fig:Ellipse_Spec} Spectra of the elliptic cavity laser
  with a cavity area $S = 8748$ $\mu$m$^2$ and an aspect ratio $p =
  0.5$ for various current values. 
The threshold current is approximately 34 mA.
}
\end{figure}

\begin{figure}
\begin{center}
  \begin{tabular}{c}
\hspace*{-0.4cm}
\raisebox{0.0cm}{\includegraphics[width=8.8cm]{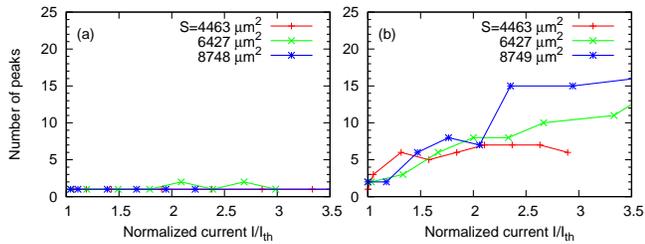}}
  \end{tabular}
\end{center}
\vspace{-4mm}
\caption{\label{fig:Peak_Size}
(Color online)
Injection current dependence of the number of peaks in the spectra of 
(a) the stadium cavity lasers and (b) the elliptic cavity lasers with  
various cavity areas $S$. In both (a) and (b), the aspect ratio is $p$ = 0.5. 
}
\end{figure}

Figure \ref{fig:Ellipse_Spec} shows the lasing spectra of the elliptic
cavity laser for various current values above $I_{th} \approx$ 34 mA 
\cite{Threshold}. 
The area and aspect ratio of the elliptic cavity are same as those of 
the stadium cavity. 
As compared to Fig. \ref{fig:Stadium_Spec}, 
there is a remarkable difference in the appearance of multiple peaks.
In order to quantify the difference of the spectral characteristics between
the two lasers, we counted the number of lasing peaks whose
intensities were larger than $-20$ dB of the maximum peak intensity in each
spectrum. 
Figure \ref{fig:Peak_Size} shows the number of peaks as
a function of the current normalized by the threshold value $I_{th}$ 
for each laser. 
As shown in Fig. \ref{fig:Peak_Size}, the difference
in the number of peaks between the two lasers increases
as $I$ increases.
We also investigated the spectral characteristics for various cavity
areas with a fixed $p$-value and found similar tendencies toward
a single-mode lasing state for the stadium cavity lasers and
toward a multimode lasing state for elliptic cavity lasers.
As a slight exception, we observed two peaks for the stadium cavity
laser 
for $I/I_{th}$ $=$ 2.1
and 2.7 in Fig. \ref{fig:Peak_Size}(a).
We attribute this to mode hopping caused by a thermal effect of the
current injection, such as a gain shift and a change in the refractive index.

\begin{figure}
\begin{center}
  \begin{tabular}{c}
\hspace*{-0.4cm}
\raisebox{0.0cm}{\includegraphics[width=8.8cm]{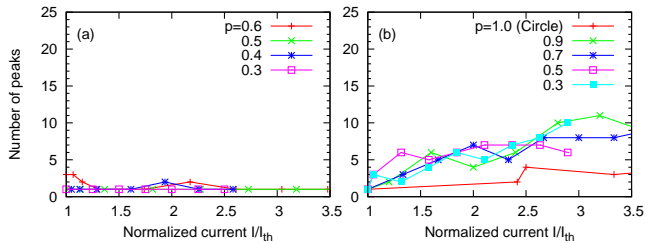}}
  \end{tabular}
\end{center}
\vspace{-4mm}
\caption{\label{fig:Peak_Deform}
(Color online)
Injection current dependence of the number of peaks in the spectra of 
(a) the stadium cavity lasers and (b) the elliptic cavity lasers with various
aspect ratios $p$. In (a), the radius $R$ is fixed as 25 $\mu$m, (i.e., the
 area for $p=0.5$ is $S$ = 4463 $\mu$m$^2$). In (b), the areas of the elliptic 
cavities are the same as that of the stadium cavity with $p=0.5$. 
}
\end{figure}

Even when changing the aspect ratios of the cavities, we found 
similar spectral characteristics. 
Figure \ref{fig:Peak_Deform}(a) shows
the current dependence of the number of peaks for the stadium
cavity lasers with various $p$-values, where the radius $R$ is fixed
at 25 $\mu$m. 
The results for the elliptic cavity lasers with various
$p$-values are shown in Fig. \ref{fig:Peak_Deform}(b).
$p=1$ corresponds to a circular cavity, which is
categorized as a non-chaotic cavity. 
There is a possibility that the number
of peaks has been underestimated because some modes have such weak
outputs (due to high confinement) that they cannot be captured in the
far field.
With a deformation toward the stadium shape (i.e., $p<1$), the
number of peaks decreases and single-mode lasing 
can be achieved for high current values 
[Fig. \ref{fig:Peak_Deform}(a)]. 
On the other hand, a deformation toward an ellipse does
not exhibit a transition to single-mode lasing
and maintains a multimode lasing state [Fig. \ref{fig:Peak_Deform}(b)].

\begin{figure}
\begin{center}
  \begin{tabular}{c}
\hspace*{-0.4cm}
\raisebox{0.0cm}{\includegraphics[width=8cm]{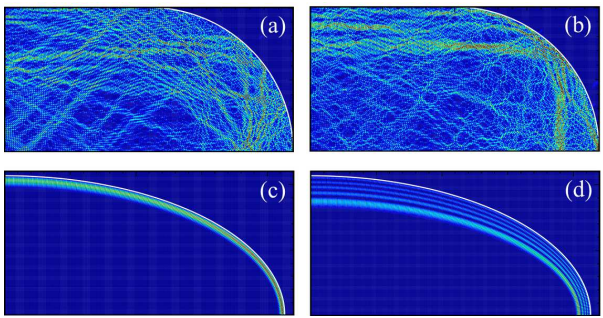}}
  \end{tabular}
\end{center}
\vspace{-4mm}
\caption{\label{fig:Wavefunc} 
(Color online) (a)(b) Spatial intensity patterns of low-loss modes 
in a quarter of the stadium cavity, 
where (b) is a scarred mode with localization along a rectangular ray orbit.
(c)(d) Spatial intensity patterns of whispering-gallery modes 
in a quarter of the elliptic cavity, 
where the radial mode number is $n_r=1$ for (c) 
and $n_r=5$ for (d).  }
\end{figure}

The above results indicate that strong suppression of multimode lasing is
a common feature of stadium cavity lasers.
As discussed in Refs. \cite{LaserPhysics,SargentIII1993}, 
a competitive interaction 
occurs among modes that are overlapped not only spectrally 
but also spatially.
In a fully chaotic cavity, modes typically have complex spatial
patterns that
spread throughout the entire cavity due to the ray dynamical
property [for a typical example, see
  Fig. \ref{fig:Wavefunc}(a)], 
and therefore result in large spatial overlaps
with other modes.
Actually, previous numerical simulations of stadium cavity lasers
demonstrated a strong
selection of lasing modes owing to a competitive interaction 
\cite{Sunada2013,Sunada2005,Harayama2005}.
Moreover,
a frequency-locking phenomenon could occur and integrate 
several lasing modes into a single lasing mode 
\cite{Harayama2003-2,Sunada2005,Sunada2013}.
On the other hand, non-chaotic cavities
typically support spatially localized modes, e.g., whispering-gallery
modes in circular and elliptic cavities.
When modes are spatially localized in different areas, 
as shown in Figs. \ref{fig:Wavefunc}(c) and \ref{fig:Wavefunc}(d), 
and the spatial overlap is small, the competition among different modes 
can be avoided.
Indeed, the simultaneous lasing of multiple whispering-gallery modes for
a circular cavity laser was numerically demonstrated in
Ref. \cite{Sunada2007}.

\begin{figure}
\begin{center}
  \begin{tabular}{c}
\hspace*{-0.4cm}
\raisebox{0.0cm}{\includegraphics[width=8cm]{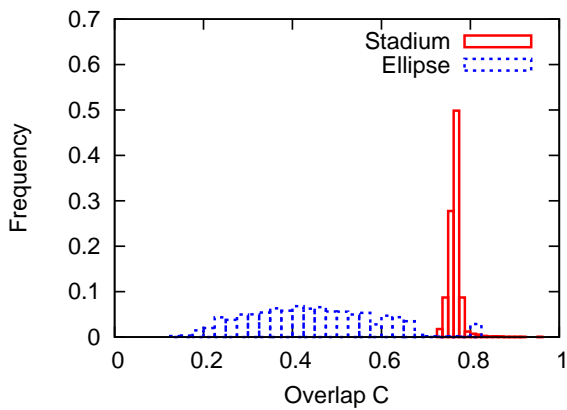}}
  \end{tabular}
\end{center}
\vspace{-4mm}
\caption{\label{fig:frequency} (Color online) Histogram of overlap
  $C$ defined by Eq. (\ref{eq:overlap}) between two low-loss modes
  for the stadium cavity (solid) and elliptic cavity (dotted).}
\end{figure}

The above discussion suggests that spatial modal overlapping is a
necessary condition for modal interactions that suppress
simultaneous lasing. 
To quantify the spatial overlaps between two modes, we introduce the
following cross-correlation for the amplitude distributions of resonance
modes:
\begin{equation}
C = 
\dfrac{
\int 
\left|
\phi(\rr)
\right|
\left|
\psi(\rr)
\right|
w(\rr)d\rr
}
{
\sqrt{
\left(
\int
\left|
\phi(\rr)
\right|^2
w(\rr)
d\rr
\right)
\left(
\int
\left|
\psi(\rr)
\right|^2
w(\rr)
d\rr
\right)
}
},
\label{eq:overlap}
\end{equation}
where $\phi(\rr)$ and $\psi(\rr)$ are the modal wave functions 
and $w(\rr)$ represents a pumping region. 
For uniform pumping, 
$w(\rr)=1$ inside the cavity, whereas $w(\rr)=0$ outside.
The correlation is essentially
similar to a spatial contribution to the cross-gain saturation 
(i.e., intensity cross-correlation) between two modes 
\cite{LaserPhysics,SargentIII1993,Yamada}.
Using the boundary element method \cite{Wiersig2003}, we calculated
the resonances of the stadium and elliptic cavities with $p$ = 0.5,
imposing a refractive index of 3.3 inside the cavities and transverse
electric (TE) polarization.  
Because of computational power limitations, 
we set a size parameter $2\pi R/\lambda \approx 100$,
where $R$ and $\lambda$ are the characteristic radius and wavelength,
respectively.  
This size parameter value is smaller than that of a real
laser cavity used in the experiments but is sufficiently 
large to discuss the properties of the wave functions 
in the short-wavelength regimes
\cite{Shinohara2008}.  
We obtained approximately 100 low-loss modes with
a quality factor $Q$ $\geq$ $3000$ for the stadium cavity,
whereas $Q$ $\geq$ $3\times 10^5$ for the elliptic cavity.  
Figure \ref{fig:frequency} shows the histogram of the spatial overlap $C$
between two low-loss modes for the two cavities.  
The $C$-values for the stadium cavity are distributed around 0.77.  
It is known that several low-loss modes in chaotic cavities are so-called
scarred modes \cite{Harayama2003_Scar,Fang2007} 
whose intensities are localized along unstable
periodic ray orbits [e.g., see Fig. \ref{fig:Wavefunc}(b)].  
An interesting finding is that there is a large overlap ($C$ $\approx$
$0.73$) even for scarred modes.
This can be explained by the fact that the localization of the scarred modes
is generally weak and significant intensities are spread throughout the
cavity. 

In contrast, the $C$-values for the elliptic cavity are widely
distributed with a mean value of 0.45. 
The relatively low $C$-values come from the small spatial overlaps
between the localized modes. 
In particular, the overlaps are small between two modes with different
radial mode numbers $n_r$, which characterize the number of field maxima
in the radial direction.  
For instance, the $C$-value between the modes with $n_r =$ 1
[Fig. \ref{fig:Wavefunc}(c)] and $n_r = $ 5 [Fig. \ref{fig:Wavefunc}(d)] 
is only 0.14. 
As seen in Fig. \ref{fig:Ellipse_Spec}, 
the lasing peaks in the spectra of the elliptic cavity laser
are not always equally spaced. 
This result means that modes with different $n_r$-values were involved
in the lasing, which supports our interpretation of the relation between 
the spatial overlaps and the spectral characteristics. 

Lastly, we emphasize that the difference in the spectral characteristics
between the stadium and elliptic cavity lasers was observed under the cw operating condition.
Under this condition, single-mode lasing has also been 
reported for other chaotic cavities with different shapes and
different gain materials \cite{Audet2007,MWKim2012}.
However, under pulsed operating conditions, multimode emission can be
observed even for fully chaotic cavity lasers \cite{Shinohara2008,Lebental2006,Choi2008,Harayama2003_Scar,Fang2007,Redding2015}.
For example, in Ref. \cite{Sunada2013}, multimode lasing has been
observed in a stadium cavity laser for a pumping pulse width of
$\lessapprox 100$ $\mu$s (see Fig. 11 of Ref. \cite{Sunada2013} for
details). 
The result can be partially attributed to transient slow mode-dynamics
towards a steady lasing state \cite{Sunada2013}.

In summary, we experimentally investigated the
difference in the spectral characteristics between
fully chaotic cavity lasers with a stadium shape and non-chaotic
cavity lasers with an elliptic shape under the cw operating condition.
In the stadium cavity lasers, 
only a single mode was excited at high pumping regimes 
regardless of the size and aspect ratio, 
whereas many modes were excited in the elliptic cavity
lasers.
The strong suppression of multimode lasing observed in the stadium cavity
lasers can be explained by the large spatial overlaps 
among the low-loss modes. 
Because it is a common feature of fully chaotic cavities that 
the modal pattern spreads throughout the cavities, 
we expect that the modal suppression leading to single-mode lasing is a universal feature of
fully chaotic cavity lasers.

\begin{acknowledgements}
The authors would like to gratefully acknowledge Professor A. D.
Stone and Professor H. Cao for fruitful discussions and insightful comments.
This work was partially supported by JSPS KAKENHI Grant No. 26790056.
\end{acknowledgements}


\begin{thebibliography}{99}
\bibitem{Threshold}
The threshold current for the elliptic cavity laser was lower than that
for the stadium cavity laser with the same area and aspect
ratio. This can be explained by the fact that the elliptic cavity
has high-Q whispering gallery modes 
(see Fig. \ref{fig:Wavefunc}).
\end{thebibliography}
\end{document}